\documentclass[conference, 10pt]{IEEEtran}
\pdfoutput=1 
\ifCLASSINFOpdf
\else
\fi
\usepackage{cite}
\usepackage{amsmath,graphicx,amssymb,color,textcomp,amsfonts,subfigure,enumerate,bm}
\usepackage{extarrows}
\usepackage{makecell,multirow,diagbox,stfloats}
\newcommand{\reffig}[1]{Fig. \ref{#1}}

\usepackage{amsthm}
\usepackage{algorithmicx,algorithm,balance, gensymb}
\usepackage{algpseudocode}
\usepackage{colortbl}
\usepackage{array}
\usepackage{booktabs}
\usepackage{tabularx}
\usepackage{forloop}
\newcounter{ct}


\begin{document}
\title{Non-Orthogonal AFDM: A Promising Spectrum-Efficient Waveform for 6G High-Mobility Communications}
\author{
	Yu Zhang$^\dag$$^*$, Qin Yi$^\dag$, Leila Musavian$^\dag$, Tongyang Xu$^*$, and Zilong Liu$^\dag$\\ 

	\normalsize 
     $^\dag$School of Computer Science and Electronics Engineering, University of Essex, CO4 3SQ Colchester, United Kingdom\\
    $^*$Department of Electronic and Electrical Engineering, University College London, London WC1E 6BT, United Kingdom\\
   Email: yu-zhang@ucl.ac.uk, \{qy24618, zilong.liu, leila.musavian\}@essex.ac.uk, tongyang.xu.11@ucl.ac.uk

}
\maketitle
\begin{abstract}
This paper proposes a spectrum-efficient non-orthogonal affine frequency division multiplexing (AFDM) waveform for reliable high-mobility communications in the upcoming sixth-generation (6G) mobile systems. Our core idea is to introduce a compression factor to enable controllable subcarrier overlapping in chirp-based AFDM modulation. To mitigate inter-carrier interference (ICI), we introduce linear precoding at the transmitter and an iterative detection scheme at the receiver. Simulation results demonstrate that these techniques can effectively reduce interference and maintain robust bit error rate (BER) performance even under aggressive compression factors and high-mobility channel conditions. The proposed non-orthogonal AFDM waveform offers a promising solution for next-generation wireless networks, balancing spectrum efficiency and Doppler resilience in highly dynamic environments.
\end{abstract}


\section{Introduction}
The upcoming sixth-generation (6G) mobile systems must cope with highly dynamic communication environments, such as high-speed trains, vehicle-to-everything (V2X) networks, unmanned aerial vehicles (UAVs), and low-earth-orbit (LEO) satellite systems. The relevant wireless channels often experience significant Doppler shifts and time-frequency selectivity. In these scenarios, traditional orthogonal frequency division multiplexing (OFDM) may be incompetent due to the high ICI incurred by large Doppler. Against this background, a novel waveform called affine frequency division multiplexing (AFDM) has been proposed recently \cite{AFDM,AFDM2}. AFDM exploits the chirp-assisted time-frequency spreading in the discrete affine Fourier transform (DAFT) domain to provide strong Doppler resilience, making it a promising waveform for reliable information exchanges in high-mobility channels.

In parallel with the need for strong Doppler resilience, the 6G wireless systems also demand high spectrum efficiency (SE) to support massive connectivity and high data rates. To improve SE, spectrally efficient frequency division multiplexing (SEFDM) \cite{SEFDM1,SEFDM2,SEFDM3} has been studied by relaxing the orthogonality constraint in multicarrier transmission. Specifically, a higher SE can be achieved by choosing a subcarrier spacing smaller than that for an OFDM system (i.e., subcarrier spacing compression). However, such compression inherently introduces ICI and therefore, sophisticated detection techniques are needed. 

A plethora of research attempts have been made for AFDM and non-orthogonal communication systems. For instance, AFDM has been integrated with sparse code multiple access (AFDM-SCMA) \cite{AFDM-SCMA}, index modulation \cite{indexmodulation}, and generalized spatial modulation \cite{GSM-AFDM} for high SE transmission. Furthermore, by fully exploiting the DAFT structure, advanced AFDM channel estimation techniques have been developed \cite{channelestimation1,channelestimation2}. Additionally, as shown in the receiver design for faster-than-Nyquist (FTN) signaling aided sparse code multiple access~\cite{FTN}, iterative detection can effectively cancel residual ICI for non-orthogonal systems.

In this paper, we propose a novel spectrum-efficient non-orthogonal AFDM waveform by advocating a compressed subcarrier spacing as that of SEFDM. By introducing a bandwidth compression factor into the AFDM modulation process, our proposed non-orthogonal waveform is able to achieve a higher SE while retaining AFDM’s resilience to high-mobility environments. Our design includes both linear precoding at the transmitter and iterative detection at the receiver, thus effectively mitigating the increased interference in non-orthogonal transmission. The key contributions of this paper are as follows:
\begin{itemize}
    \item  \textbf{Non-Orthogonal AFDM Waveform Design:}  To the best of our knowledge, this is the first study to introduce a non-orthogonal AFDM waveform. By exploiting the benefits of non-orthogonal signals, our design improves SE and is well-suited for bandwidth-limited and high-mobility scenarios.
    \item \textbf{Transmitter Precoding:} We employ both zero forcing (ZF) and minimum mean squared error (MMSE) linear precoding schemes to pre-compensate for channel distortions and suppress ICI induced by subcarrier overlap. By carefully shaping the transmitted signal based on channel state information, the precoding stage ensures that the effective channel experienced by the receiver is nearly diagonal, thereby simplifying the subsequent detection process. 
    \item \textbf{Iterative Detection:} At the receiver, we develop an iterative detection (ID) algorithm for a non-orthogonal AFDM system. The proposed detection algorithm iteratively cancels residual interference and refines the estimates of the transmitted symbols, leading to improved bit error rate (BER) performance.
\end{itemize} 
 
The rest of this paper is organized as follows. In Section~\ref{Sec:SystemModel}, we present the system model for the proposed non-orthogonal AFDM scheme, including the channel model and signal structure. Section~\ref{Sec:Design} outlines the proposed transmitter-side precoding strategy, while Section~\ref{Sec:Detection} details the iterative detection algorithm at the receiver side. We then provide simulation results in Section~\ref{Sec:Results}. Finally, we conclude the paper in Section~\ref{Sec:Conclusion} with a summary of our findings and discussions on future research directions.

\section{System Model}\label{Sec:SystemModel}
In this section, we present an overview of the AFDM waveform, followed by the proposed non-orthogonal AFDM waveform that integrates SEFDM. The goal is to preserve AFDM’s robustness to Doppler spreads while benefiting from SEFDM’s improved SE.
\subsection{Waveform of AFDM}
The AFDM transmit signal for a block of $N$ symbols can be expressed as \cite{AFDM2}
\begin{align}\label{eqAFDM}
    s[n]=\frac{1}{\sqrt{N}}\sum\limits_{m=0}^{N-1} x[m] e^{\imath2\pi(c_1n^2+c_2m^2+\frac{nm}{N})},  n=0, ..., N-1,
\end{align}
where $x[m]$ is the information symbol, and $c_1$, $c_2$ are the AFDM time and frequency domain chirp parameters, respectively. The terms $c_1n^2$ and $c_2m^2$ introduce a quadratic phase modulation in both the time and frequency domains.
The matrix form of \eqref{eqAFDM} is given by
\begin{align}
    \mathbf{s}=\mathbf{\Lambda}_{c_1}^H\mathbf{F}^H\mathbf{\Lambda}_{c_2}^H\mathbf{x},
\end{align}
where $\mathbf{\Lambda}_c=\text{diag}(e^{\imath2\pi cn^2}, n=0, 1, ..., N-1)$, $\mathbf{x}$ denote the vector of information symbols, and $\mathbf{F}$ is the discrete Fourier transform (DFT) matrix with entries $e^{-\imath2\pi mn/N}/\sqrt{N}$.

Similar to cyclic prefix (CP) in OFDM, AFDM employs a chirp-periodic prefix (CPP) to mitigate inter-block interference arising from the multipath fading. The CPP is defined as
\begin{align}
    s[n]=s[N+n]e^{-\imath2\pi c_1(N^2+2Nn)}, \quad n=-L_{\text{cp}}, ..., -1,
    \end{align}
    where $L_{\text{cp}}$ is the length of CPP.
    
After passing through a time-varying multipath channel, the received samples $r[n]$ can be expressed as 
\begin{align}
    r[n]=\sum\limits_{l=0}^\infty s[n-l]g_n(l)+w[n],
\end{align}
where $w[n]\sim \mathcal{CN}(0, N_0)$ is the additive Gaussian noise and $g_n(l)$ is the impulse response of channel at time $n$ and delay $l$. The channel may exhibit both delay and Doppler spreads in practical high-mobility environments.

In AFDM, demodulation is performed via the inverse operation of \eqref{eqAFDM}, which projects the received signal $r[n]$ back into the DAFT domain, as follows
\begin{align}\label{rxAFDM}
    y[m]=\frac{1}{\sqrt{N}}\sum\limits_{n=0}^{N-1} r[n] e^{-\imath2\pi(c_1n^2+c_2m^2+\frac{nm}{N})}.
\end{align}
Let $\mathbf{r}$ represent the vector of received time-domain samples, and $\mathbf{y}$ denote the vector of demodulated (DAFT-domain) symbols. Then \eqref{rxAFDM} can be written in matrix form as
\begin{align}
\mathbf{y}=\mathbf{\Lambda}_{c_2}\mathbf{F}\mathbf{\Lambda}_{c_1}\mathbf{r}=\mathbf{Ar}=\mathbf{H}_{\text{eff}}\mathbf{x}+\tilde{\mathbf{w}},
\end{align}
where $\mathbf{A} = \mathbf{\Lambda}_{c_2} \mathbf{F} \mathbf{\Lambda}_{c_1}$ is a unitary matrix, $\tilde{\mathbf{w}}=\mathbf{Aw}$ is the transformed noise vector, and $\mathbf{H}_{\text{eff}}=\mathbf{AH}\mathbf{A}^H$ represents the effective channel in the DAFT domain with 
$\mathbf{H}$ being the channel matrix that captures the time-frequency dispersion of the physical propagation channel.

\subsection{Waveform of Non-Orthogonal AFDM}

\begin{figure}[t!]
\begin{center}
\includegraphics[width=90mm]{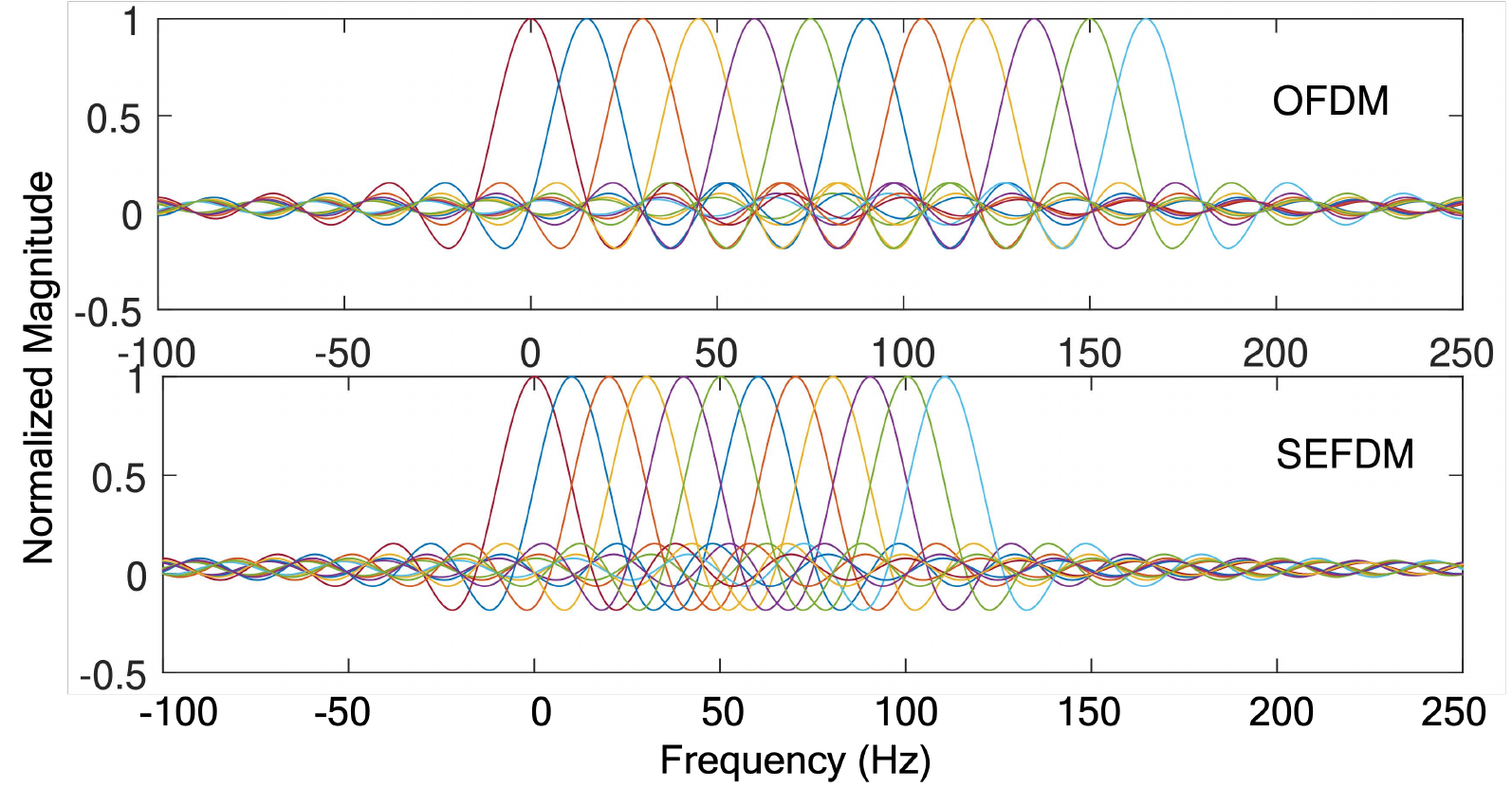}
\end{center}
\caption{Frequency-domain comparison of OFDM and SEFDM subcarriers.}
\label{Fig:SEFDM}
\end{figure}
\begin{figure*}[h!]
    \centering
    \includegraphics[width=140mm]{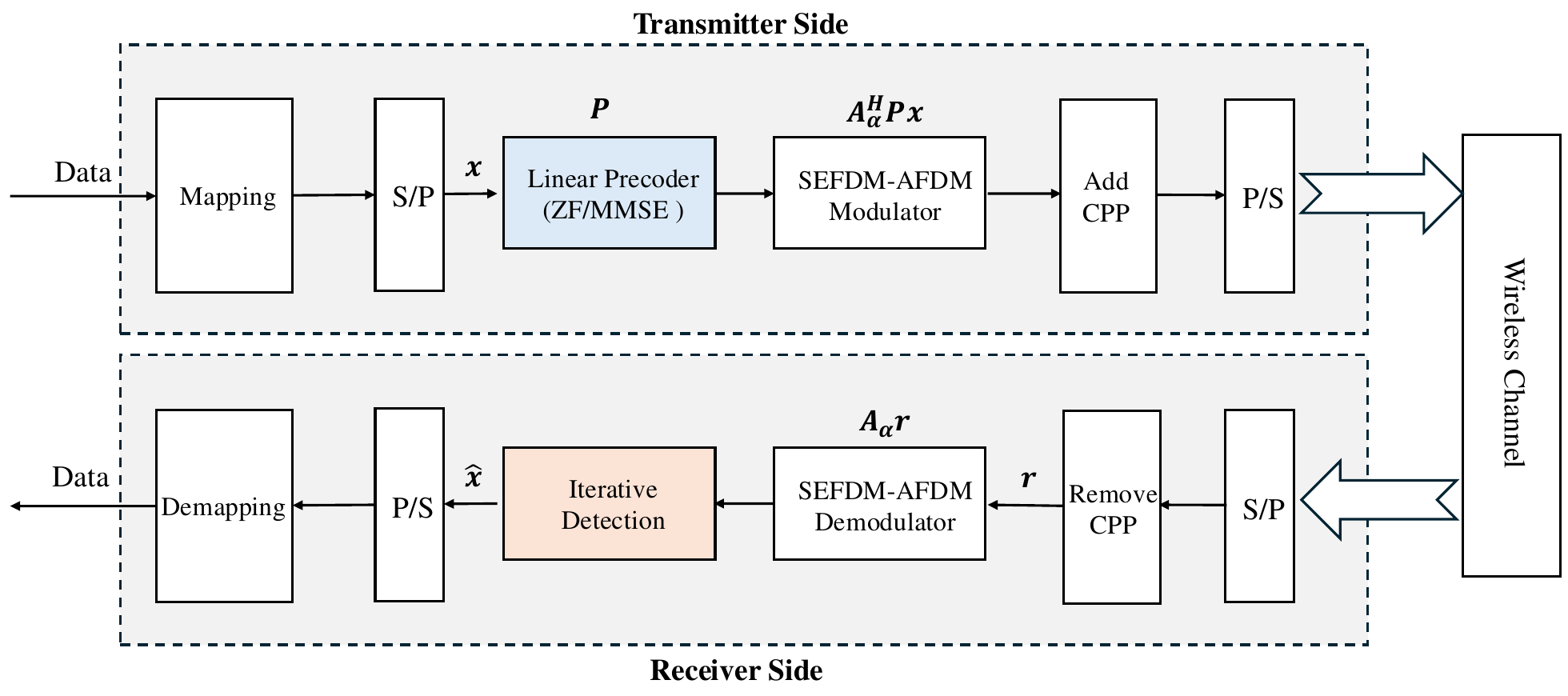}
    \caption{The diagram of the transmitter and receiver design for the non-orthogonal AFDM waveform.}
    \label{fig:framework}
\end{figure*}
To further improve SE while retaining the resilience of AFDM to time-varying channels, we consider a hybrid waveform that combines SEFDM \cite{SEFDM1} with AFDM. 
The continuous-time SEFDM signal uses $N$ subcarriers with subcarrier spacing $\Delta f$ over one symbol duration $T$. It is given by 
\begin{align}\label{xt}
x(t)=\frac{1}{\sqrt{T}}\sum\limits_{l=-\infty}^{\infty}\sum\limits_{n=0}^{N-1}s_{l,n}\exp\left[\frac{j2\pi n\alpha(t-lT)}{T}\right],
\end{align}
where $\alpha = \Delta f T$ denotes the bandwidth compression factor, $0<\alpha<1$, and $s_{l, n}$ is the symbol  (e.g., QAM or QPSK) modulated on the $n$-th subcarrier of the  $l$-th SEFDM symbol. When $\alpha=1$, the SEFDM symbol becomes a standard OFDM symbol, meaning there is no bandwidth compression. The bandwidth saving percentage value is $(1-\alpha)\times 100\%$, as shown in \reffig{Fig:SEFDM}. Decreasing the value of $\alpha$ can provide greater bandwidth savings at the cost of increased subcarrier overlap.

For practical implementation, the continuous-time SEFDM signal $x(t)$ is sampled at intervals $T/Q$ to generate the discrete SEFDM signals, as follows
\begin{align}\label{SEFDM symbol}
X[k]=\frac{1}{\sqrt{Q}}\sum\limits_{n=0}^{N-1}s_n\exp\left(\frac{j2\pi \alpha k n}{Q}\right),
\end{align}
where $Q=\rho N$ denotes the number of time samples and $\rho\geq 1$ is the oversampling factor. This formulation illustrates how SEFDM achieves spectral compression by scaling the subcarrier spacing with the factor $\alpha$.

By integrating SEFDM into the AFDM framework, we can benefit from the bandwidth savings of SEFDM as well as the Doppler resilience by AFDM signals.
The combined AFDM and SEFDM signal can be expressed as
\begin{align}\label{eqAFDM_SEFDM}
    s^{(\alpha)}[n] = \frac{1}{\sqrt{N}} \sum\limits_{m=0}^{N-1} x[m] e^{\imath 2\pi \left( c_1 n^2 + c_2 m^2 + \frac{\alpha nm}{N} \right)},
\end{align}
where $\alpha$ is the SEFDM compression factor, $0<\alpha < 1$.

The waveform in \eqref{eqAFDM_SEFDM} can also be represented in matrix form as
\begin{align}\label{symbol}
    \mathbf{s}^{(\alpha)} = \mathbf{\Lambda}_{c_1}^H \mathbf{F}_\alpha^H \mathbf{\Lambda}_{c_2}^H \mathbf{x}=\mathbf{A}_\alpha^H\mathbf{x},
\end{align}
where $\mathbf{x} = [x[0], x[1], \ldots, x[N-1]]^T$ is the input symbol vector, $\mathbf{\Lambda}_{c_1} = \text{diag}\left(e^{-\imath 2\pi c_1 n^2}, \, n = 0, \ldots, N-1 \right)$ is the AFDM chirp modulation matrix in the time domain, $\mathbf{\Lambda}_{c_2} = \text{diag}\left(e^{-\imath 2\pi c_2 m^2}, \, m = 0, \ldots, N-1 \right)$ is the AFDM chirp modulation matrix in the frequency domain,
$\mathbf{A}_{\alpha} = \mathbf{\Lambda}_{c_2} \mathbf{F}_\alpha \mathbf{\Lambda}_{c_1}$ is the combined modulation matrix for AFDM and SEFDM, and $\mathbf{A}_{\alpha}(m,n) = e^{-\imath 2\pi \left( c_1 n^2 + c_2 m^2 + \frac{\alpha nm}{N} \right)} $, $\mathbf{F}_\alpha$ is the SEFDM discrete Fourier transform (DFT) matrix with elements $ F_{\alpha, m, n} = \frac{1}{\sqrt{N}} e^{-\imath 2\pi \frac{\alpha mn}{N}}, m, n = 0, \ldots, N-1$.

Similar to the demodulated AFDM signal in \eqref{rxAFDM}, the received samples of non-orthogonal AFDM are demodulated as
\begin{align}\label{rxSEFDM-AFDM}
    y_{\alpha}[m] &= \frac{1}{\sqrt{N}} \sum\limits_{n=0}^{N-1} r[n] e^{-\imath 2\pi \left( c_1 n^2 + c_2 m^2 + \frac{\alpha nm}{ N} \right)}.
\end{align}
Then \eqref{rxSEFDM-AFDM} can be written in matrix form as
\begin{align}
\mathbf{y}_{\alpha}=\mathbf{\Lambda}_{c_2}\mathbf{F}_{\alpha}\mathbf{\Lambda}_{c_1}\mathbf{r}=\mathbf{A}_{\alpha}\mathbf{r}=\mathbf{H}_{\text{eff},\alpha}\mathbf{x}+\tilde{\mathbf{w}},
\end{align}
where $\mathbf{H}_{\text{eff},\alpha} = \mathbf{A}_{\alpha} \mathbf{H} \mathbf{A}_{\alpha}^H$ is the effective channel matrix, and $\tilde{\mathbf{w}} = \mathbf{A}_{\alpha} \mathbf{w}$ is the transformed noise vector.

The non-orthogonal AFDM signal can employ the advantage of SEFDM's high spectral efficiency and AFDM's Doppler resilience, which provides a potential solution for time-variant channels with limited bandwidth scenarios.

\section{Transmitter Design for Non-orthogonal AFDM}\label{Sec:Design}
In non-orthogonal AFDM, the receiver must handle both channel distortions and inter-carrier interference (ICI) caused by the reduced subcarrier spacing. To reduce receiver complexity and improve detection performance, we introduce a linear precoder at the transmitter. The goal is to pre-compensate for the effective channel seen by the non-orthogonal AFDM waveform. In particular, we consider two popular linear precoding schemes, i.e., zero forcing (ZF) and minimum mean squared error (MMSE), as shown in \reffig{fig:framework}.

In \eqref{symbol}, the transmit signal without precoding is $ \mathbf{s}^{(\alpha)}=\mathbf{A}_\alpha^H\mathbf{x}$. With linear precoding, the signal is transmitted with precoding matrix $\mathbf{P} \in \mathbb{C}^{N \times N}$, as follows
\begin{equation}
    \mathbf{s}_{\text{pre}} = \mathbf{P} \mathbf{A}_\alpha^H \mathbf{x}.
\end{equation}

After passing through an $N \times N$ channel matrix $\mathbf{H}$ and demodulated by $\mathbf{A}_\alpha$ at the receiver, the system model becomes
\begin{equation}
    \mathbf{y}_{\text{pre}} = \mathbf{A}_\alpha \mathbf{H}  \mathbf{P}\mathbf{A}_\alpha^H  \mathbf{x} + \mathbf{A}_\alpha  \mathbf{w},
\end{equation}
where $\mathbf{y}_{\text{pre}}$ is the non-orthogonal AFDM-domain received signal and $\mathbf{w}$ is the additive noise vector.

\subsubsection{ZF Precoding}
Zero forcing precoding aims to eliminate all channel distortions, ignoring noise enhancement, as follows
\begin{equation}
    \mathbf{H} \mathbf{P}_\mathrm{ZF} \approx \mathbf{I}.
\end{equation}
If the matrix $\mathbf{H}$ is invertible, the exact ZF solution is given by \cite{precoding}
\begin{equation}
    \mathbf{P}_\mathrm{ZF} =\left(\mathbf{H}\right)^\dagger= \mathbf{H}^H\left(\mathbf{H}\mathbf{H}^H\right)^{-1},
\end{equation}
where $(\cdot)^\dagger$ denotes the Moore--Penrose pseudo-inverse. Under perfect channel state information (CSI), this solution ideally cancels channel-induced distortion. However, noise and subcarrier overlapping can be amplified, potentially degrading performance at low signal-to-noise ratios (SNRs).

\subsubsection{MMSE Precoding}
In scenarios where noise enhancement is a concern, MMSE precoding provides a balanced approach by minimizing $\mathbb{E} \bigl\|\mathbf{x} -\mathbf{A}_\alpha \mathbf{H}  \mathbf{P}\mathbf{A}_\alpha^H  \mathbf{x}\bigr\|^2$,
yielding the matrix
\begin{equation}
    \mathbf{P}_\mathrm{MMSE}=\mathbf{H}^H\left(\mathbf{H}\mathbf{H}^H+\sigma^2\mathbf{I}\right)^{-1},
\end{equation}
where $\sigma^2$ denotes the noise variance. Unlike ZF, MMSE precoding avoids large gains on weak eigenmodes of the channel, thus offering robust performance across a range of SNRs.

\subsubsection{Implementation Considerations}
To implement either ZF or MMSE precoding, the transmitter requires accurate CSI and knowledge of the non-orthogonal AFDM modulation matrix $\mathbf{A}_\alpha$. After computing the precoding matrix $\mathbf{P}$, the precoded time-domain samples are generated by $\mathbf{s}_{\text{pre}} = \mathbf{A}_\alpha^H \mathbf{P}\mathbf{x}$. A chirp-periodic prefix (CPP) is then appended before converting the digital samples to analog for over-the-air transmission. The detailed process of the transmitter side design is shown in the diagram of \reffig{fig:framework}.

Precoding effectively reshapes the transmitted signal so that the overall channel modulation product approaches an identity-like mapping between $\mathbf{x}$ and $\mathbf{y}$. Consequently, the receiver can perform simpler detection or reduce the number of iterations required for inter-carrier interference (ICI) cancellation, leading to enhanced robustness under both multipath propagation and Doppler effects.

\section{Iterative Detection for Non-orthogonal AFDM}\label{Sec:Detection}
In AFDM, the frequency spacing $\Delta f$ between overlapping chirps satisfies $\Delta f = \frac{1}{T}$ to ensure the mutual independence of subcarriers, where $T$ is the OFDM symbol period. However, in non-orthogonal AFDM, the subcarriers are tightly arranged in a non-orthogonal manner. This results in inter-carrier interference (ICI) between subcarriers, which poses a key challenge for the non-orthogonal AFDM system.

\begin{algorithm}\label{ID detection}
	\caption{Non-Orthogonal AFDM Iterative Detection Algorithm}
	\begin{algorithmic}[1]
		\State \textbf{Input:} received signal $y_{\text{eq}}$, modulation matrix $A_{\alpha}$, the number of iterations $N_{\text{iter}}$, Modulation scheme $M$ (e.g., `QPSK', `16-QAM'), Parameters for soft detection (e.g., thresholds \(r_1\) for QPSK, and \(T_1, T_2\) for 16-QAM).
		\State \textbf{Output:} final detected symbols $\hat{\mathbf{x}}$.
		
		\State  \textbf{Step 1: Initialization}
	\State  Compute the initial symbol estimates: $\mathbf{x}^{(0)} = y_{\text{eq}}$
		\State  Compute the modulation correlation matrix: $\mathbf{A}_{\alpha}^{\text{corr}} = A_{\alpha} A_{\alpha}^H$
		\State Remove self-interference: $\mathbf{A}_{\alpha}^{\text{corr}}(i, i) = 0, \forall i$
		
		\State  \textbf{Step 2: Iterative Refinement}
		\For{$k = 1$ to $N_{\text{iter}}$}
		\State  Compute interference: $\mathbf{I}^{(k)} = \mathbf{A}_{\alpha}^{\text{corr}} \mathbf{x}^{(k-1)}$
	\State  Cancel interference: $\mathbf{r}^{(k)} = y_{\text{eq}} - \mathbf{I}^{(k)}$
    \If{\(M\) equals QPSK}
        \For{\(n = 1\) to length\((\mathbf{r}^{(k)})\)}
            \State Let \(r_n \gets \mathbf{r}^{(k)}(n)\)
            \State \(\Re\{x^{(k)}_n\} \gets \begin{cases}
                  +1, & \Re\{r_n\} > r_1, \\
                  -1, & \Re\{r_n\} < -r_1, \\
                  \Re\{r_n\}, & \text{otherwise}
               \end{cases}\)
            \State \(\Im\{x^{(k)}_n\} \gets \begin{cases}
                  +1, & \Im\{r_n\} > r_1, \\
                  -1, & \Im\{r_n\} < -r_1, \\
                  \Im\{r_n\}, & \text{otherwise}
               \end{cases}\)
        \EndFor
    \ElsIf{\(M\) equals 16-QAM}
        \For{\(n = 1\) to length\((\mathbf{r}^{(k)})\)}
            \State Let \(r_n \gets \mathbf{r}^{(k)}(n)\)
            \State \(\Re\{x^{(k)}_n\} \gets \begin{cases}
                  +3, & \Re\{r_n\} > T_1, \\
                  +1, & T_2 < \Re\{r_n\} \le T_1, \\
                  -1, & -T_1 \le \Re\{r_n\} < -T_2, \\
                  -3, & \Re\{r_n\} < -T_1, \\
                  \Re\{r_n\}, & \text{otherwise}
               \end{cases}\)
            \State \(\Im\{x^{(k)}_n\} \gets \begin{cases}
                  +3, & \Im\{r_n\} > T_1, \\
                  +1, & T_2 < \Im\{r_n\} \le T_1, \\
                  -1, & -T_1 \le \Im\{r_n\} < -T_2, \\
                  -3, & \Im\{r_n\} < -T_1, \\
                  \Im\{r_n\}, & \text{otherwise}
               \end{cases}\)
        \EndFor
    \Else
        \State \(\mathbf{x}^{(k)} \gets \textsc{SoftDetect}(\mathbf{r}^{(k)}, M)\)
        \Comment{Use a generic soft detection function for other modulations.}
    \EndIf
\EndFor
	\end{algorithmic}
\end{algorithm}

To eliminate the ICI in non-orthogonal AFDM signals, we propose an iterative detection algorithm, as shown in Algorithm 1.
We first do the MMSE channel equalization of the received signal $\mathbf{r} = \mathbf{H} \mathbf{s} + \mathbf{w}$ to mitigate the effects of the channel matrix $\mathbf{H}$, which is computed as
\begin{align}
\mathbf{r}_{\text{eq}} = \left( \mathbf{H}^H \mathbf{H} + \sigma^2 \mathbf{I} \right)^{-1} \mathbf{H}^H \mathbf{r},
\end{align}
where $\sigma^2$ is the noise variance and $\mathbf{I}$ is the identity matrix. Then we transform the equalized signal to the non-orthogonal AFDM domain, as follows
    \begin{align}
        \mathbf{y}_{\text{eq}} = \mathbf{A}_{\alpha} \mathbf{r}_{\text{eq}}.
    \end{align}

If we assume the interference arises only from the overlapping subcarriers due to SEFDM compression, then it can be modeled using the modulation correlation matrix:
\begin{align}
    \mathbf{A}_{\alpha}^{\text{corr}} = \mathbf{A}_{\alpha} \mathbf{A}_{\alpha}^H.
\end{align}
This correlation matrix captures both the self-interference (diagonal elements) and inter-carrier interference (off-diagonal elements). Analyzing the correlation matrix for interference cancellation can avoid explicit modeling of the physical channel matrix $\mathbf{H}$.
To eliminate the self-interference, we can set the diagonal elements of $\mathbf{A}_{\alpha}^{\text{corr}}$ to zero, given by
\begin{align}
    \mathbf{A}_{\alpha}^{\text{corr}}(i, i) = 0, \quad \forall i.
\end{align}

For the inter-carrier interference, we need to use the iteration algorithm to cancel it. Let $\mathbf{x}^{(k)}$ be the estimate of the transmitted symbols at iteration \(k\), then the ICI is approximated by
    \begin{align} \label{ICI}
        \mathbf{I}^{(k)} = \mathbf{A}_{\alpha}^{\text{corr}} \mathbf{x}^{(k-1)}.
    \end{align}
By subtracting the interference \eqref{ICI} from the received signal, we can get the interference cancellation, as follows
    \begin{align}
        \mathbf{r}^{(k)} = \mathbf{y}_{\text{pre}} - \mathbf{I}^{(k)}.
    \end{align}
Then perform the soft or hard detection step on \(\mathbf{r}^{(k)}\) to update \(\mathbf{x}^{(k)}\). 
After $N_{\text{iter}}$ iterations, the output the final detected symbols is
\begin{align}
    \hat{\mathbf{x}} = \mathbf{x}^{(N_{\text{iter}})}.
\end{align}

\section{Simulation Results}\label{Sec:Results}
In this section, we present numerical evaluations of the bit error rate (BER) performance for the proposed non-orthogonal AFDM system under various compression factors $\alpha$, detection strategies, and precoding schemes. The number of propagation paths is set to 3, and the Doppler shift for each path is computed using Jakes' formula, i.e.,  $\nu_i = \nu_{\text{max}} \cos(\theta_i)$, where $\nu_{\text{max}}$ denotes the maximum normalized Doppler shift, and $\theta_i$ is uniformly distributed within $[-\pi, \pi]$. The system employs 32 subcarriers at a carrier frequency of 4~GHz, with a subcarrier spacing of 1~kHz, resulting in a total bandwidth of 32~kHz. To capture high-mobility conditions, the maximum speed is set to 540~km/h, which corresponds to a maximum Doppler shift of 2~kHz. Additionally, a maximum delay spread of 2 symbol durations is considered to ensure frequency-selective fading. Finally, 4QAM modulation is used throughout the simulations to clearly demonstrate the performance trends of the AFDM waveform.
 
\begin{table}[h!]
\centering
\caption{SIMULATIONS SETTING}
\label{tab:simulation_settings}
\begin{tabular}{|l|l|}
\hline
\textbf{Parameter} & \textbf{Value} \\ \hline
The number of subcarriers, \( N \) & 32 \\ \hline
Carrier frequency, \( f_c \) & 4 GHz \\ \hline
AFDM subcarrier spacing, \( \Delta f\) & 1 kHz \\ \hline
Bandwidth & 32 kHz \\ \hline
Maximum speed & 540 km/h \\ \hline
Maximum Doppler shift & 2 kHz \\ \hline
Samples of the maximum
delay spread of channel & 2 \\ \hline
Modulation scheme for communication & 4QAM \\ \hline
The number of iterations for ID & 20 \\ \hline
\end{tabular}
\end{table}

\begin{figure}
    \centering
    \includegraphics[width=95mm]{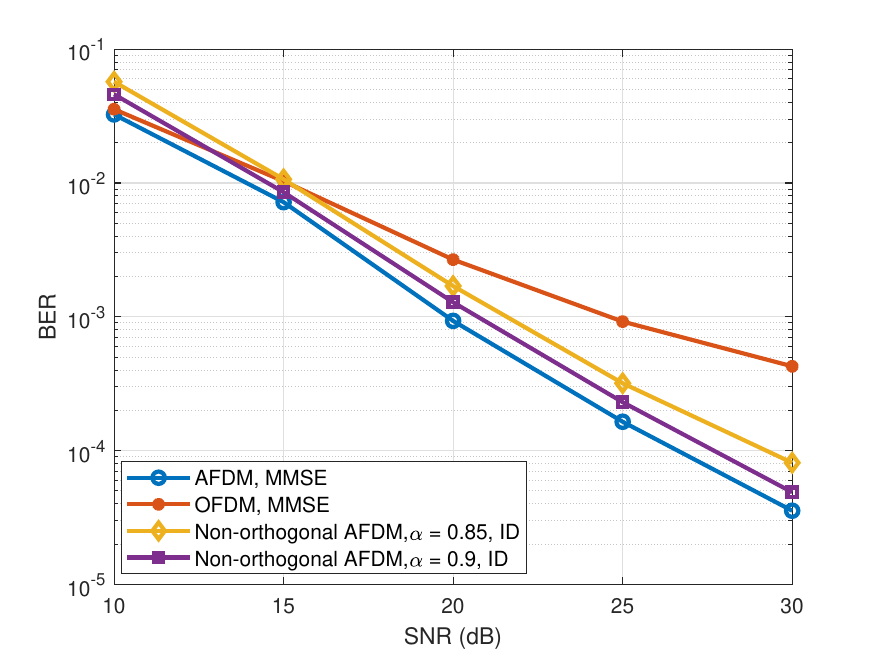}
    \caption{BER for AFDM and OFDM with MMSE detection and non-orthogonal AFDM with iterative detection. $\alpha=0.85$ (spectral efficiency improvement is 17.65\%). $\alpha=0.9$ (spectral efficiency improvement is 11.11\%).}
    \label{fig:NonAFDM_OFDM}
\end{figure}

\reffig{fig:NonAFDM_OFDM} shows the BER performance of non-orthogonal AFDM, AFDM, and OFDM. Compared to OFDM, non-orthogonal AFDM demonstrates superior BER performance. Therefore, under the same bandwidth conditions, non-orthogonal AFDM can achieve simultaneous improvements in both BER performance and spectral efficiency compared to OFDM, which provides a promising approach for bandwidth-limited and high-mobility communication scenarios.

\begin{figure}
    \centering
    \includegraphics[width=95mm]{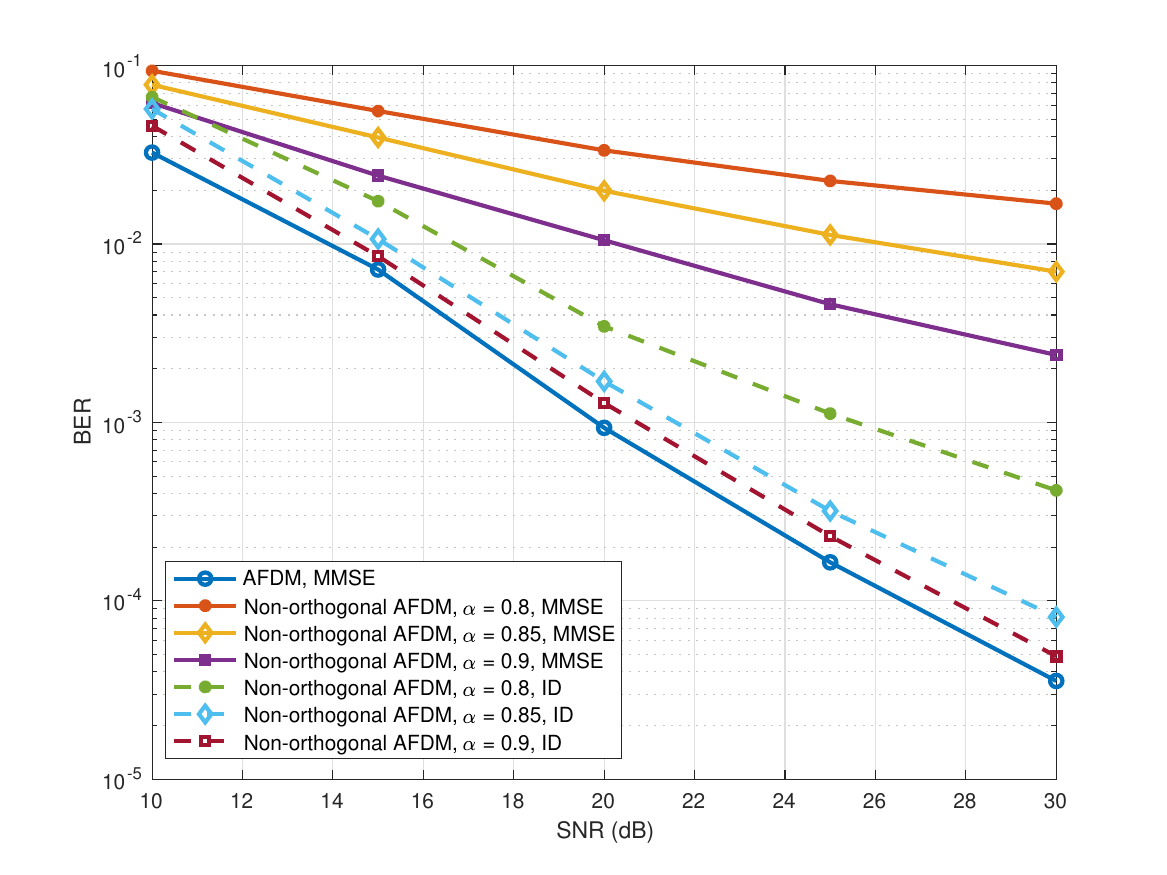}
    \caption{BER performance comparison for MMSE and iterative detection in non-orthogonal AFDM with compression factors $\alpha=0.8, 0.85, 0.9$.}
    \label{fig:ID}
\end{figure}

\reffig{fig:ID} compares the BER performance of the non-orthogonal AFDM system under two different detection strategies, i.e., MMSE detection and iterative detection. The SEFDM compression factors are set as $\alpha\in\{0.8,0.85,0.9\}$. As the SNR increases, all curves exhibit a downward trend of the BER. For both MMSE and ID methods, the BER performs worse when the $\alpha$ decreases. This is because lower $\alpha$ suffers from higher interference due to increased subcarrier overlap. However, the ID can partially mitigate this interference compared to the MMSE method. For $\alpha=0.9$, the performance of ID method comes very close to that of the standard orthogonal AFDM signal. This phenomenon indicates that iterative processing can effectively handle the increased interference at lower compression factors.

\begin{figure}
    \centering
    \includegraphics[width=95mm]{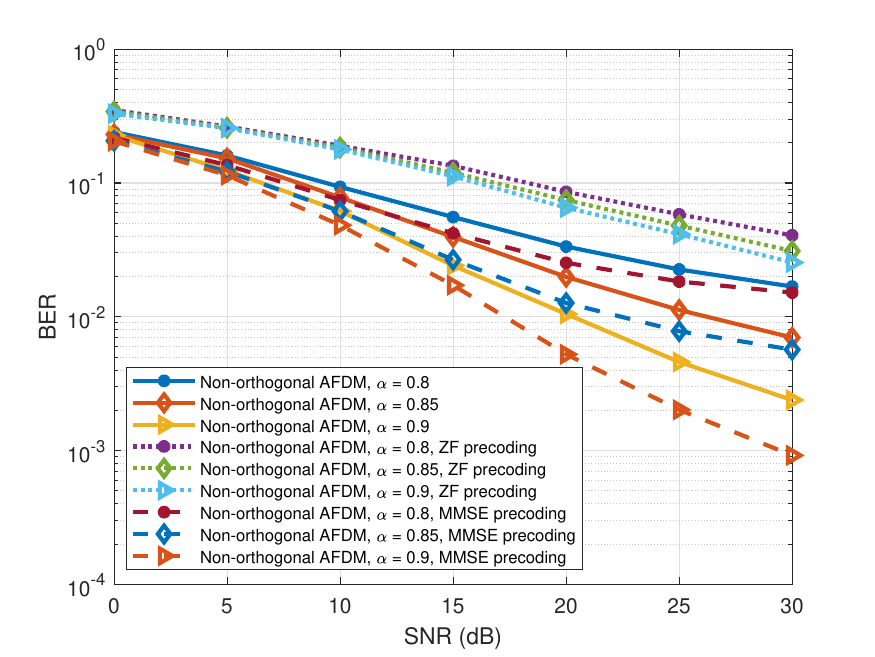}
    \caption{BER performance of non-orthogonal AFDM for different precoding methods ($\alpha=0.8, 0.85, 0.9$).}
    \label{fig:MMSE}
\end{figure}

\reffig{fig:MMSE} illustrates the BER performance of the non-orthogonal AFDM system when ZF and MMSE precoding are employed at the transmitter. In this scenario, precoding is designed to pre-cancel subcarrier interference introduced by the non-orthogonal subcarriers. As expected, MMSE precoding curves show a noticeable performance improvement compared to the unprecoded non-orthogonal AFDM signal, particularly at moderate SNR values. This is because MMSE precoding effectively balances interference cancellation with noise enhancement, making it more robust than ZF.  The BER performance of MMSE precoding is better than that of ZF precoding. On the other hand, ZF precoding fails to improve the BER performance compared to the non-precoded scheme. This is due to the fact that in ill-conditioned channels or when certain eigenmodes of the channel are weak, ZF will apply very large gains on those weak modes, significantly amplifying the noise. Consequently, MMSE precoding consistently outperforms ZF in terms of BER performance.



\balance

\section{Conclusion}\label{Sec:Conclusion}
In this work, we have presented a novel non-orthogonal AFDM waveform by integrating SEFDM-based bandwidth compression into the AFDM framework. To mitigate the ICI introduced by the non-orthogonal transmission, we have studied linear precoding at the transmitter, including ZF and MMSE precoding. Additionally, we have developed an iterative detection algorithm at the receiver for refined signal detection and hence enhanced BER performance. It is shown that the proposed iterative detection algorithm can effectively mitigate the ICI, while the proposed MMSE precoding offers a good trade-off between interference suppression and noise enhancement. Future research includes new detection algorithms for further interference mitigation and the exploration of advanced coding techniques to improve reliability.

\section{Acknowledgement}
This work was supported by the UK Engineering and Physical Sciences Research Council (EPSRC) under Grants EP/X035352/1 (`DRIVE'), EP/Y000986/1 (`SORT'), EP/X04047X/1 (`TITAN'), EP/Y037243/1 (`TITAN Extension'), and EP/Y000315/2.

\end{document}